\documentclass[prb,twocolumn,showpacs,preprintnumbers,amsmath,noamssymb,superscriptaddress]{revtex4}

\usepackage{graphicx}
\usepackage{dcolumn}
\usepackage{bm}

\newcommand{\ybco}{YBa$_2$Cu$_3$O$_{6+x}$}
\newcommand{\tbco}{Tl$_2$Ba$_2$CuO$_6$}
\newcommand{\tbcod}{Tl$_2$Ba$_2$CuO$_{6+\delta}$}
\newcommand{\bscco}{Bi$_2$Sr$_2$CaCu$_2$O$_{8+x}$}
\newcommand{\tc}{$T_\textrm{c}$}
\newcommand{\rs}{$R_\textrm{s}$}
\newcommand{\xs}{$X_\textrm{s}$}

\begin{document}


\title{Electrical transport measurements in the superconducting state of \\
\bscco\ and \tbcod}
 
\author{S.~\"Ozcan}
\affiliation{Cavendish Laboratory, University of Cambridge, Madingley Road,
Cambridge, CB3 0HE, United Kingdom}
\author{P.~J.~Turner}
\affiliation{Department of Physics, Simon Fraser University,\\
Burnaby, British Columbia, V5A~1S6, Canada}
\author{J.~R.~Waldram}
\affiliation{Cavendish Laboratory, University of Cambridge, Madingley Road,
Cambridge, CB3 0HE, United Kingdom}

\author{R.~J.~Drost}
\author{P.~H.~Kes}
\affiliation{Kamerlingh Onnes Laboratory, Leiden University, P.O. Box
9504, 2300 RA Leiden, The Netherlands}
\author{D.~M.~Broun}
 \email{dbroun@sfu.ca}
\affiliation{Department of Physics, Simon Fraser University,\\
Burnaby, British Columbia, V5A~1S6, Canada}

\date{\today}

\begin{abstract}
Precise measurements of the in-plane microwave surface impedance of high-quality single crystals of \bscco\ and \tbcod\ are used to probe the
relaxation time of nodal quasiparticles in the \mbox{$d$-wave} superconducting state through a two-fluid analysis of the microwave conductivity. While this analysis requires us to posit a form for the frequency-dependent quasiparticle conductivity, we clearly demonstrate that the extraction of the relaxation rate is quite insensitive to the assumed shape of the quasiparticle spectrum.  The robustness of the analysis is rooted in the oscillator-strength sum rule and the fact that we simultaneously measure the real \textit{and} imaginary parts of the conductivity.   In both \bscco\ and \tbcod\ we infer a linear temperature dependence of the transport relaxation rate $1/\tau$ and a small but finite zero-temperature intercept.  The linear temperature dependence of $1/\tau$ is in accord with expectations for weak \textit{elastic} scattering in an unconventional superconductor with line nodes and a small residual density of states.   The same analysis reveals an onset of inelastic scattering at  higher temperatures similar to that seen in the \ybco\ superconductors.  Finally we extrapolate the two-fluid model over a range of frequencies up to five times the measurement frequency, where the extrapolation predicts behaviour that is qualitatively similar to terahertz conductivity data on \bscco\ thin films.
While relaxation rates in \bscco\ and \tbcod\  are substantially higher than in \ybco\  there are qualitative similarities between all three materials, and the differences can likely be attributed to varying levels of static disorder.  We therefore conclude that a universal picture of quasiparticle scattering in the cuprates is emerging.
\end{abstract}

\maketitle

\section{Introduction}

A controversial issue in the field of high temperature superconductivity is the nature of the intrinsic quasiparticle damping. Given the structural similarities of the different cuprate compounds and the similar doping dependence of many of their physical properties it is natural to expect the dominant mechanism of quasiparticle scattering to be universal.  The experimental situation, however, is currently unclear. Early terahertz \cite{nuss91} and microwave conductivity  \cite{bonn92} measurements on \ybco\   revealed a rapid drop in the quasiparticle relaxation rate upon entering the superconducting state. Photoemission \cite{valla99,valla00} and terahertz \cite{corson00} spectroscopies of \bscco\ provide a different picture, with a relaxation rate for the nodal quasiparticles that is much larger than that of \ybco, and is of the order of the thermal energy  over the full temperature range of the superconducting state.  Combined with the simple energy--temperature scaling of the single-particle lifetime observed in the photoemission experiments, this has led to proposals that proximity to a quantum critical point is responsible for the unusual quasiparticle dynamics in \bscco.\cite{vojta00}  
In order to draw a more universal picture it is important to establish to what extent the intrinsic quasiparticle relaxation in the superconducting state of \ybco\  differs from that in other important cuprate materials such as \bscco\ and \tbcod. This might also clarify the differences in the transport relaxation rate measured by the microwave conductivity and the single-particle scattering rate measured by photoemission experiments.  Experimental probes sensitive to charge dynamics that can cross this divide and be applied to both \ybco\ and \bscco\ are therefore of great importance to further progress in this area.  

Equally intriguing are the dynamical properties at the lowest temperatures, which should be dominated by nodal quasiparticles scattering from static disorder.  The effect of impurities on the transport properties of a \mbox{\mbox{$d$-wave}} superconductor is in general a complicated problem, requiring in most cases a self-consistent treatment of multiple scattering of quasiparticles from impurities.  The standard theoretical model used for these calculations is the self-consistent $t$-matrix approximation (SCTMA),\cite{hirschfeld93a,hirschfeld94,rieck99}  which treats quasiparticle scattering within the framework of \mbox{\mbox{$d$-wave}} BCS superconductivity and Fermi liquid theory.  In this model two simple regimes of scattering exist. The first is the weak-scattering or \textit{Born}  limit, in which the energy-dependent relaxation rate $1/\tau(E)$ is determined by the total phase space for recoil. As a result, $1/\tau(E)$ should be proportional to the density of states $N(E)$, which in a clean \mbox{\mbox{$d$-wave}} superconductor is linear in energy.  In that case the thermally averaged relaxation rate $1/\tau(T)$ should have a linear temperature dependence.   Recent broadband microwave spectroscopy of Ortho-II-ordered YBa$_2$Cu$_3$O$_{6.52}$ has for the first time revealed clear signatures of Born-like scattering, including a linear temperature dependence of the average relaxation rate and conductivity spectra with upwards curvature down to the lowest frequencies.\cite{turner03}  The other regime in which the SCTMA makes simple predictions is in the unitarity limit, in which the impurity has a quasiparticle bound state at the Fermi energy that acts as a zero-energy scattering resonance, with $1/\tau(E) \propto 1/E$. Calculations have shown unitarity-limit scatterers to be strongly pair-breaking\cite{hirschfeld93a,hirschfeld94}  and have been strikingly confirmed by experiments in which the controlled addition of Zn impurities drives a crossover from linear to quadratic temperature dependence of the London penetration depth $\lambda_\textrm{L}$.\cite{bonn94a}  More recent measurements of the frequency-dependent microwave conductivity of high purity Ortho-I YBa$_2$Cu$_3$O$_{6.993}$ have refined the experimental situation and presented a puzzle --- a temperature-independent relaxation rate below 20~K.\cite{hosseini99}  This is compatible with neither the Born nor unitarity limits and has prompted extensions of the SCTMA theory to include the effect of `order-parameter holes', local suppressions of the pairing amplitude at the impurity site that act as an additional scattering channel.\cite{hettler00}  In the last year the SCTMA has been further extended to take into account small-angle scattering.\cite{nunner05} This latest work is extremely promising, as the theory now starts from realistic models of the disorder potential, including that due to impurities located \textit{out of} the CuO$_2$ planes.  Among its successes is the ability to model a range of experiments, including microwave and thermal conductivity, angle-resolved photoemission and scanning tunneling spectroscopy, using just \textit{one} set of parameters to describe the concentration, location and scattering phase shifts of impurities.\cite{scalapino05}

In this paper we use electrical transport experiments at microwave frequencies to probe the quasiparticle lifetime in near-optimally doped single crystals of \bscco\  and \tbcod.   Both the microwave conductivity $\sigma = \sigma_1 - \textrm{i} \sigma_2$ and the d.c.\ resistivity are sensitive to the same fundamental physical property --- the quasiparticle transport relaxation rate. However, unlike d.c.\ resistivity experiments, measurements at high frequencies can probe electrical transport deep within the superconducting state.  Since we simultaneously measure both the real and imaginary parts of the microwave conductivity we are well positioned to extract the temperature-dependent relaxation rate using a two-fluid analysis.  This has been a somewhat speculative endeavor in the past, with only limited experimental data available to support the assumption of a Lorentzian conductivity spectrum.\cite{bonn92,bonn93,bonn94,waldram97} The two-fluid description of the conductivity we adopt here is based on a phenomenological conductivity spectrum $\sigma_1(\omega)$ that has been carefully validated  through detailed spectroscopic measurements on \ybco.\cite{turner03,harris05}  Nevertheless, we take care to show that the qualitative form of the temperature-dependent relaxation rate inferred from this analysis is quite insensitive to the assumed form of the conductivity spectrum.  That this is so is primarily due to constraints imposed by the oscillator strength sum rule.
In this paper we report new data on \bscco\ at temperatures down to 2~K, which we analyze along with previously published data on \bscco\  and \tbcod.\cite{bscco,tbco}  The high resolution of the new experiments allows the scattering dynamics of the nodal quasiparticles to be revealed in great detail. In  \bscco\  and \tbcod\ the dominant sources of disorder are known to be located away from CuO$_2$ planes and should therefore act as weak, small-angle scatterers.  We find that the transport relaxation rates in  \bscco\  and \tbcod\  have a similar temperature dependence, with $1/\tau(T) \propto T$ at low temperature, as expected for a \mbox{\mbox{$d$-wave}} superconductor in the weak scattering limit.   This is similar  to observations made in  Ortho-II YBa$_2$Cu$_3$O$_{6.52}$,\cite{turner03} although the absolute relaxation rates are substantially higher in \bscco\ and \tbco.   At higher temperatures $1/\tau(T)$ increases more strongly with temperature, also reminiscent of transport in the \ybco\ superconductors.\cite{hosseini99}  Our main conclusion is that while relaxation rates in \bscco\ and \tbcod\ are higher than in \ybco, there are qualitative similarities, with the differences most likely due to variations in the level of static disorder.  

\section{Experimental Techniques}

\bscco\  crystals were grown in Leiden using the travelling-solvent floating-zone (TSFZ) method \cite{li94} and annealed in air at $500^{\circ}$C for 3 days to produce samples with \tc\ = 87~K.  The samples have low levels of impurities and sharp superconducting transitions. They show a high degree of crystallinity, with x-ray rocking curves and magnetic-torque measurements indicating a mosaic spread less than $0.2^{\circ}$.  In addition, the crystals were shown to be free of extended defects using magneto-optic Kerr
microscopy.  The crystal used in the microwave experiment was obtained by repeatedly cleaving a larger piece from a TSFZ boule to obtain a small
$0.5~\textrm{mm} \times 0.5~\textrm{mm} \times 0.02~\textrm{mm}$ platelet with shiny, mirror-like surfaces and sharp edges.  In order to test and extend the
generality of the conclusions we draw from this new data we have carried out the same analysis on previously published data from
TSFZ \bscco\  crystals  grown in Tsukuba by Kadowaki and Mochiku\cite{bscco, mochiku93} and flux-grown \tbcod\  crystals prepared in Cambridge by Mackenzie and Tyler.\cite{tbco,liu92}  Structural disorder in these materials has been characterised by electron-probe
microanalysis (EPMA), which reveals cation nonstoichiometry at about the 10\% level, an intrinsic consequence of thermal-equilibrium fluctuations in composition at the growth temperature.  The EPMA measurements give average compositions of Tl$_{1.85}$Ba$_2$Cu$_{1.15}$O$_{6+\delta}$ for the Tl-cuprate, indicating Cu
substitution onto the Tl sites, and Bi$_{2.16}$Sr$_{1.91}$Ca$_{1.03}$Cu$_2$O$_{8+x}$ for the Bi-cuprate. Since this cation disorder is located away
from the CuO$_2$ planes the in-plane charge dynamics will only be weakly influenced by the impurity potentials, which should give rise to
predominantly small-angle scattering.\cite{varma01}

\begin{figure}[t]
\includegraphics[width = 85mm]{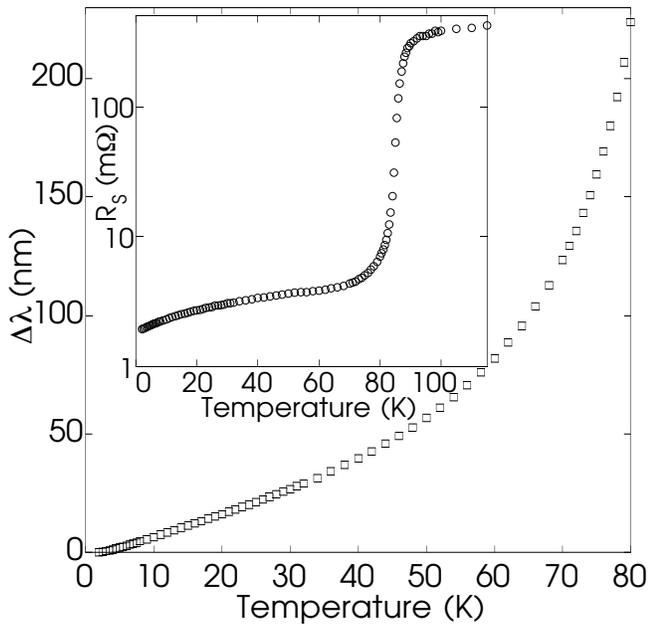}
\caption{\label{lambdars} Penetration depth data for the Leiden \bscco\
crystal show $\Delta\lambda(T) \propto T$ at low temperature.  Inset: 38.6~GHz
surface resistance data for the same crystal shows a sharp superconducting transition at $T_\textrm{c} = 87$~K and a low residual surface resistance of 2~m$\Omega$.}
\end{figure}

\begin{figure}[t]
\includegraphics[width = 85mm]{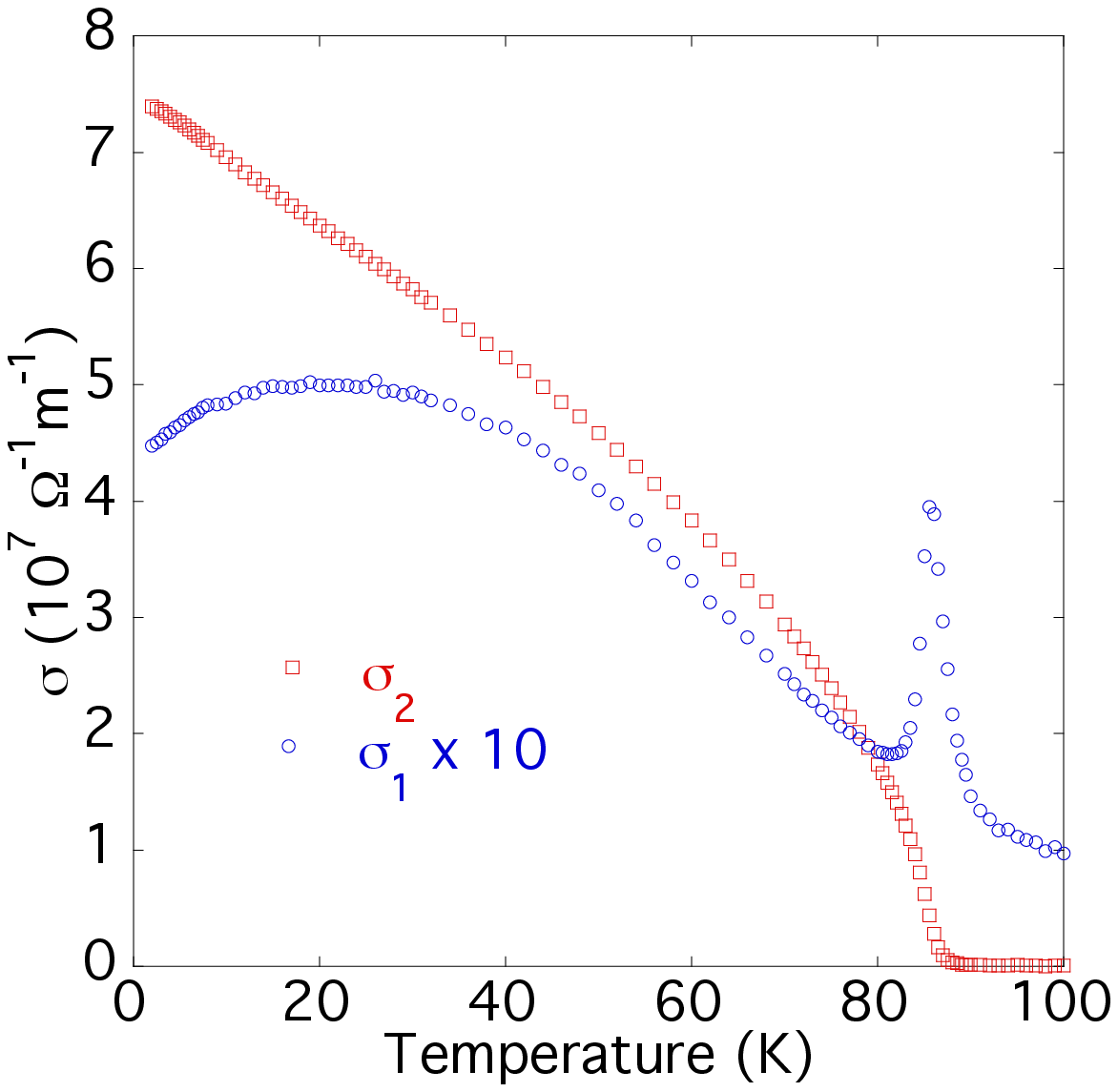}
\caption{(Color online) The 38.6~GHz microwave conductivity of the Leiden \bscco\ crystal inferred from the surface impedance data in Fig.~\ref{lambdars}. $\sigma_1(T)$ is scaled up by a factor of 10 for clarity.  A characteristic feature of all microwave conductivity data on \bscco\ and \tbcod\ is the finite zero-temperature intercept of $\sigma_1(T)$.}
\label{conduct}
\end{figure}

Measurements of the surface impedance $Z_\textrm{s} = R_\textrm{s} + \textrm{i} X_\textrm{s}$ were made in circular cylindrical
TE$_{011}$ Nb cavity resonators, held at the helium bath temperature (1.2~K or 4.2~K) and supporting TE$_{011}$ modes with $Q$ factors in the range $2
\times 10^6$ to $3 \times 10^7$. During these measurements the sample is positioned inside the cavity on a movable and independently-heated sapphire rod at an
antinode of the microwave magnetic field, which is oriented perpendicular to the $a$-$b$ plane of the sample.  Although this geometry involves substantial demagnetising 
fields for our platelet samples, it results in primarily in-plane screening currents being induced, an important requirement for highly anisotropic materials like \bscco\  and
\tbcod.  The surface impedance of the sample is obtained from the measured cavity response using the cavity perturbation formula\cite{ormeno97}
\begin{equation}
\Delta f_\textrm{B}(T) - 2\textrm{i}\Delta f_0(T) = \Gamma(R_\textrm{s} + \textrm{i} \Delta X_\textrm{s}),
\end{equation}
where $\Delta
f_\textrm{B}(T)$ is the change in bandwidth of the TE$_{011}$ mode upon inserting the sample into the cavity,  $\Delta f_0(T)$ is the shift in resonant frequency upon warming the sample from $T_{\textrm{base}}$ to $T$, and $\Gamma$ is an empirically-determined scale factor that depends only on the geometry 
of the sample and cavity.  A key feature of our method is that  \rs$(T)$ and $\Delta$\xs$(T) \approx \omega \mu_0 \Delta\lambda(T)$ are measured \textit{at the same time and on the same sample}, allowing both $\sigma_1(T)$ and $\sigma_2(T)$ to be determined at the measurement frequency in a model-independent way.

\section{Experimental Results}  
  
Figure~\ref{lambdars} shows \rs$(T)$ and $\Delta\lambda(T) = \lambda(T) - \lambda(2~\textrm{K})$ for the Leiden \bscco\ crystal at 38.6~GHz. 
At \tc\ $ = 87$~K \rs$(T)$ has a sharp superconducting transition, indicating good sample homogeneity, then decreases monotonically below
\tc\ to a low residual value of 2~m$\Omega$.  No discernible change in \rs\ was observed as the input power of the resonator was
varied over more than two orders of magnitude implying an absence of extended defects, which are expected to give nonlinearities associated with weak superconducting links..  At low temperatures $\Delta\lambda(T)$ has 
a strong, linear temperature dependence, a hallmark of \mbox{\mbox{$d$-wave}} pairing in the cuprates \cite{hardy93} only observed in samples with low levels of strong-scattering defects.\cite{bonn94a}

The complex microwave conductivity is obtained from \rs\ and $\Delta$\xs\ using the local-limit expression $\sigma_1 - \textrm{i} \sigma_2 = \textrm{i} \omega \mu_0/ Z_\textrm{s}^2$ assuming, as
in earlier work,\cite{bscco} that $\lambda(T=0) = 2100$~\AA.  A different choice of $\lambda(T=0)$ does not modify the qualitative features of $\sigma(T)$,  or  of $1/\tau(T)$ obtained from the two-fluid analysis. The conductivity data are shown in Fig.~\ref{conduct}, with $\sigma_1(T)$ expanded by a factor of ten.  While similar in form to  previously published results on Tsukuba-grown \bscco\ crystals, the new data are of higher resolution and extend to lower temperature.  At low temperature $\sigma_2(T) \propto 1/\lambda^2(T)$ has a strong linear temperature dependence, a consequence of line nodes in the energy gap. 
The data near \tc\ are reminiscent of \ybco, for which the superfluid density approaches zero with vertical slope due to 3D-XY critical fluctuations.\cite{kamal94}
Fluctuations are also responsible for the sharp peak in $\sigma_1(T)$ just below \tc,\cite{waldram99} which is followed at lower temperatures by a broad peak
near 20~K before decreasing slightly to a zero-temperature intercept of $4.5 \times 10^6~\Omega^{-1}$m$^{-1}$.

\section{Two-Fluid Model}

In this section we develop a two-fluid model in order to extract the thermally averaged quasiparticle relaxation rate from the microwave conductivity data. The complex conductivity of a superconductor has two contributions that conduct in parallel: $\sigma_{s}$ from the superfluid condensate; and $\sigma_{qp}$ from quasiparticle excitations.  The superfluid conductivity is dissipationless and takes the form
\begin{equation}
\sigma_{s}(\omega)=\frac{n_\textrm{s} e^2}{m^\ast} \left(\pi \delta(\omega) + \frac{1}{\textrm{i}\omega} \right).
\label{superfluid}
\end{equation}
The zero-frequency delta function is required by causality and represents the energy absorbed in accelerating the superfluid.  The overall scale of $\sigma_\textrm{s}$ is parameterized by $n_\textrm{s}/m^\ast$,  the ratio of the effective density of superconducting electrons $n_\textrm{s}$ to their effective mass $m^\ast$.  This quantity is temperature dependent and is closely related to the London penetration depth $\lambda_\textrm{L}$:
\begin{equation}
\frac{n_\textrm{s}(T)}{m^\ast} = \frac{1}{\mu_0 e^2 \lambda_\textrm{L}^2(T)}.
\end{equation}

The quasiparticle conductivity has a more complicated form than the superfluid term.  It contains information on both the quasiparticle excitation spectrum \textit{and}  the quasiparticle charge dynamics.  In a \mbox{\mbox{$d$-wave}} superconductor the conventional approach to calculating the quasiparticle conductivity is the self-consistent $t$-matrix approximation, which accurately takes into account the multiple scattering of quasiparticles from impurities, a process that leads to pair-breaking and scattering resonances.\cite{hirschfeld93a,hirschfeld94,rieck99}  It has been shown that the SCTMA calculations of the conductivity, for \textit{any} scattering strength, can be well approximated by a simple, energy-averaged Drude form:\cite{hirschfeld94}
\begin{equation}
\sigma(\omega,T) = \frac{e^2}{m^\ast}  \int_{-\infty}^\infty N(E) \left(-\frac{\partial f(E)}{\partial E} \right) \frac{1}{\textrm{i} \omega + 1/\tau(E)} \textrm{d}E,
\label{genDrude}
\end{equation}
where $1/\tau(E)$ is the energy-dependent electrical-transport relaxation rate, $N(E)$ is the density of states of the \mbox{$d$-wave} superconductor and $f(E)$ is the Fermi function at temperature $T$.   In a \mbox{$d$-wave} superconductor the relaxation rate takes a simple form in two important limits: for weak scattering (the Born limit) $1/\tau(E) \propto E$; while for strong scattering (the unitarity limit) $1/\tau(E) \propto 1/E$, with a scattering resonance at zero energy.  The energy dependence of the relaxation rate leaves a clear signature in the quasiparticle conductivity spectrum, making microwave frequency measurements well suited to probing the quasiparticle dynamics.

Several experiments have attempted to map out the detailed form of the quasiparticle conductivity spectrum in the cuprates,\cite{hosseini99,turner03,harris05} with the clearest to date being that of Turner \textit{et al.}\ who used a novel broadband spectrometer to measure the microwave conductivity of \mbox{Ortho-I} YBa$_2$Cu$_3$O$_{6.993}$ and Ortho-II YBa$_2$Cu$_3$O$_{6.52}$ as continuous functions of frequency from 1~GHz to 21~GHz.   These measurements revealed conductivity spectra that were broadly in accord with those expected from the SCTMA calculations, with upwards curvature down to the lowest frequencies.\cite{turner03,harris05}  It was shown by Turner \textit{et al.} that the conductivity spectra were well described by a simple phenomenological form:
\begin{equation}
\sigma_1^\textrm{qp}(\omega) = \frac{\sigma_\textrm{dc}}{1 + (\omega \tau)^y}.
\label{conductivity}
\end{equation}
For $y =2$ this spectrum has the familiar Drude form of a single Lorentzian.  As $y$ is reduced below 2 the upwards curvature extends to lower frequencies and the spectrum starts to correspond well to the expected form for a \mbox{\mbox{$d$-wave}} superconductor with energy-dependent scattering.  In fits to experimental data in Ref.~\onlinecite{turner03} it was found that for Ortho-II-ordered YBa$_2$Cu$_3$O$_{6.52}$ $y \approx 1.45$ and for fully oxygenated YBa$_2$Cu$_3$O$_{6.993}$ $y \approx 1.67$.   

There are several caveats to note when using Eq.~\ref{conductivity}.  For $y \ne 2$ the spectrum applies only at  positive frequencies.  Also, $y$ must be greater than one or the spectrum cannot be integrated, in violation of the oscillator-strength sum rule.  Finally, at high frequencies $\sigma(\omega) \sim 1/\omega^y$, whereas physically we expect $\sigma(\omega)$ to eventually go as $1/\omega^2$ since in all cases the energy-dependent relaxation rate entering Eq.~\ref{genDrude} will have an upper bound.  Eq.~\ref{conductivity} can therefore only be approximate at the highest frequencies. Nevertheless, it has provided extremely good fits to the measured spectra for \ybco\ in the frequency range 1~GHz to 21~GHz.\cite{turner03}  In those experiments Eq.~\ref{conductivity} also appeared to extrapolate very well outside the measured frequency range, as the conductivity spectral weight inferred from integrating fits of Eq.~\ref{conductivity} to the \ybco\ data very accurately tracked the loss of superfluid spectral weight. Nonanalytic conductivity spectra have been proposed in several other cases to describe the unusual optical conductivity of cuprate and ruthenate materials in the normal state.\cite{ioffe98,vandermarel99,dodge00}  In that work, the form of the conductivity was
\begin{equation}
\sigma(\omega) = \frac{\sigma_\textrm{dc}}{(1/\tau + \textrm{i} \omega)^\alpha}.
\end{equation}
Spectra of this sort suffer the same limitations as Eq.~\ref{conductivity}, but in each case their use was motivated by good fits to experimental data.  With these warnings in mind we will continue on and use Eq.~\ref{conductivity} in the analysis of our conductivity data.  We will be careful to show, though, that our conclusions do not depend on the assumed shape of $\sigma(\omega)$.

With a phenomenological spectrum that we can extrapolate \textit{outside} the measured frequency range,  the imaginary part of the quasiparticle conductivity can be obtained by a Kramers--Kr\"onig transform:
\begin{equation}
\sigma^\textrm{qp} =  \frac{\sigma_\textrm{dc}}{1 + (\omega \tau)^y} - \textrm{i} \sigma_\textrm{dc} \mbox{KK}\!\left(\omega \tau\right),
\label{qpconductivity}
\end{equation}
where
\begin{equation}
\mbox{KK}\!\left(\omega \tau \right) = - \frac{2\omega}{\pi} \mathcal{P} \int_0^\infty \frac{1}{1 + \left(\omega^\prime \tau \right)^y} \frac{1}{\omega^{\prime 2} - \omega^2}\textrm{d}\omega^\prime.
\end{equation}

\begin{figure} [t]
\includegraphics[width = 85mm]{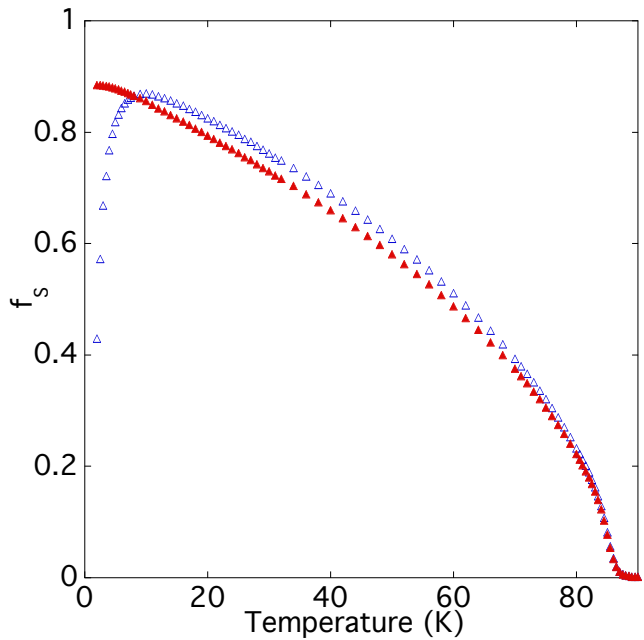}
\caption{(Color online) Superfluid fraction $f_\textrm{s}(T)$ from a two-fluid analysis ($y = 2$) of conductivity data from the Leiden \bscco\ crystal.  Open triangles show $f_\textrm{s}(T)$ for  $\sigma_0   = 7.5\times10^7~\Omega^{-1}$m$^{-1} = \sigma_2(T \rightarrow 0)$.  This choice of $\sigma_0$ leads to a unphysical downturn in $f_\textrm{s}(T)$ at low temperature.  Solid triangles shows $f_\textrm{s}(T)$ for $\sigma_0 = 8.3\times10^7~\Omega^{-1}$m$^{-1}$, a choice that allows for a small amount of residual conductivity spectral weight as $T \rightarrow 0$.  $f_\textrm{s}$ is now monotonic in temperature.}
\label{twofluid} 
\end{figure}
\begin{figure} [t]
\includegraphics[width = 85mm]{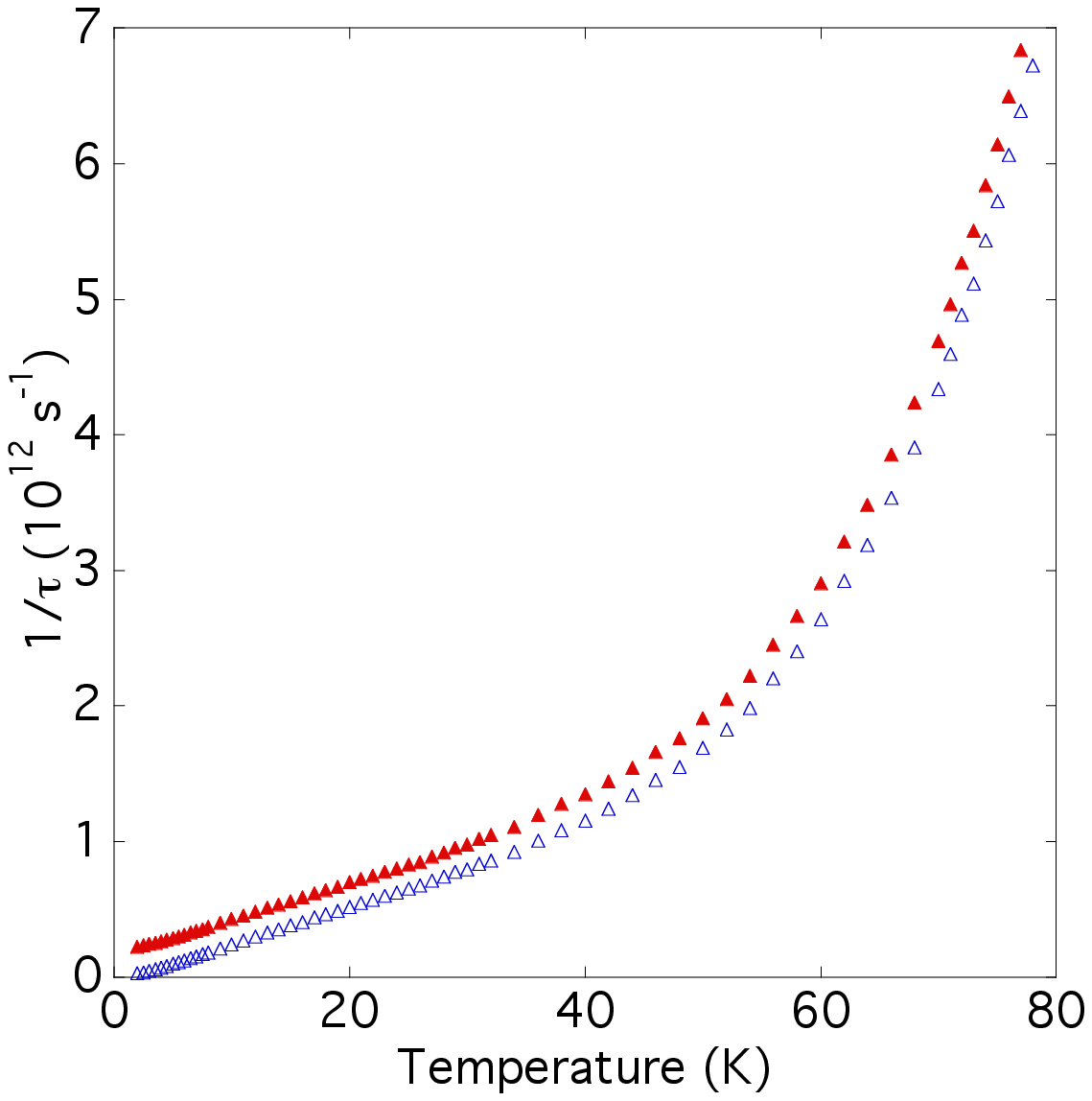}
\caption{ (Color online)  Relaxation rate $1/\tau(T)$ from a two-fluid analysis ($y = 2$) of the microwave conductivity of the Leiden \bscco\ crystal.  Open triangles correspond to $\sigma_0 = 7.5\times10^7~\Omega^{-1}$m$^{-1}$, solid triangles to $\sigma_0 = 8.3\times10^7~\Omega^{-1}$m$^{-1}$.  Allowing for residual normal fluid in the choice of $\sigma_0$ results in a small offset to $1/\tau(T)$.  The data reveal a clear linear temperature dependence of $1/\tau(T)$ at low temperature followed by a strong increase in scattering at high temperatures.}
\label{scatteringrate}
\end{figure}

Eqs.~\ref{superfluid} and \ref{qpconductivity} can be combined into a phenomenological model of the full microwave conductivity of a \mbox{$d$-wave} superconductor at finite frequencies, which we will later use to  extract the temperature dependent relaxation rate of the quasiparticles.   The model is
\begin{equation}
\sigma  = \frac{n e^2}{m^\ast} \left(\frac{f_\textrm{s}}{\textrm{i}\omega}  + \frac{f_\textrm{n} \tau y \sin (\pi/y)}{2} \left[\frac{1}{1 + \left(\omega \tau\right)^y} - \textrm{i}\;\mbox{KK}\!\left(\omega \tau \right) \right]\right).
\label{gentwofluidmodel}
\end{equation}
Here the prefactor of the quasiparticle term is obtained by insisting that the frequency integral of Eq.~\ref{conductivity}
\begin{equation}
\int_0^\infty\frac{\sigma_\textrm{dc}}{1+ (\omega\tau)^y} \textrm{d} y = \frac{\sigma_\textrm{dc}}{\tau} \frac{\pi}{y \sin(\pi/y)}
\end{equation}
be independent of $y$ and equal to $\frac{\pi}{2} f_\textrm{n} n e^2/m^\ast$, as required by the oscillator-strength sum rule.  In the clean limit, $\hbar/\tau \ll 2\Delta$, the sum rule ensures that the superfluid and quasiparticle conductivity spectral weights sum to a temperature independent value, imposing the constraint that \mbox{$f_\textrm{s} + f_\textrm{n} = 1$}.  This has been shown to apply in the case of the \ybco\ superconductors \cite{harris05} and should certainly be the case in our \bscco\ and \tbcod\ crystals, as the superconducting energy gap $\Delta$ is large and the samples clean.

In the limit $y=2$, Eq.~\ref{gentwofluidmodel} reverts to the standard Drude two-fluid model,
\begin{equation}
\sigma = \frac{ne^2}{m^\ast}\left(\frac{f_\textrm{s}}{\textrm{i}\omega} + \frac{f_\textrm{n}}{1/\tau + \textrm{i}\omega}\right),
\end{equation}
which in the past has been used  to carry out analyses broadly similar  to those presented below, albeit in the absence of experimental evidence to support the use of a Lorentzian spectrum.\cite{bonn92,bonn93,bonn94,waldram97, berlinsky93}  For $y=2$ there are closed-form expressions for the relaxation rate and normal-fluid fraction:\cite{waldram97}
\begin{eqnarray}
1/\tau & = & \omega\;\frac{\sigma_0 - \sigma_2(T)}{\sigma_1(T)}\\
f_\textrm{n} & = & (1 + \omega^2\tau^2)\left(1 - \frac{\sigma_2(T)}{\sigma_0}\right).
\end{eqnarray}
Here $\sigma_0 = ne^2/m^\ast\omega$ represents the \textit{total} conductivity spectral weight and in general may be different from $\sigma_2(T\rightarrow 0)$ due to the pair-breaking effects of disorder.  Making allowance for residual normal fluid at $T = 0$ is an important part of the analysis that follows.

For $y \ne 2$, closed-form expressions for the relaxation rate and normal fluid fraction do not exist, forcing us to invert Eq.~\ref{gentwofluidmodel} numerically.  Setting $y < 2$ leads to conductivity spectra  that are very different from the Drude form, but we show in the next section that the qualitative temperature dependence of $1/\tau$ is insensitive to the choice of $y$.  This is a consequence of the oscillator strength sum rule and helps to make our analysis very robust.

\section{Analysis and Discussion}

\begin{figure}[t]
\includegraphics[width = 85mm]{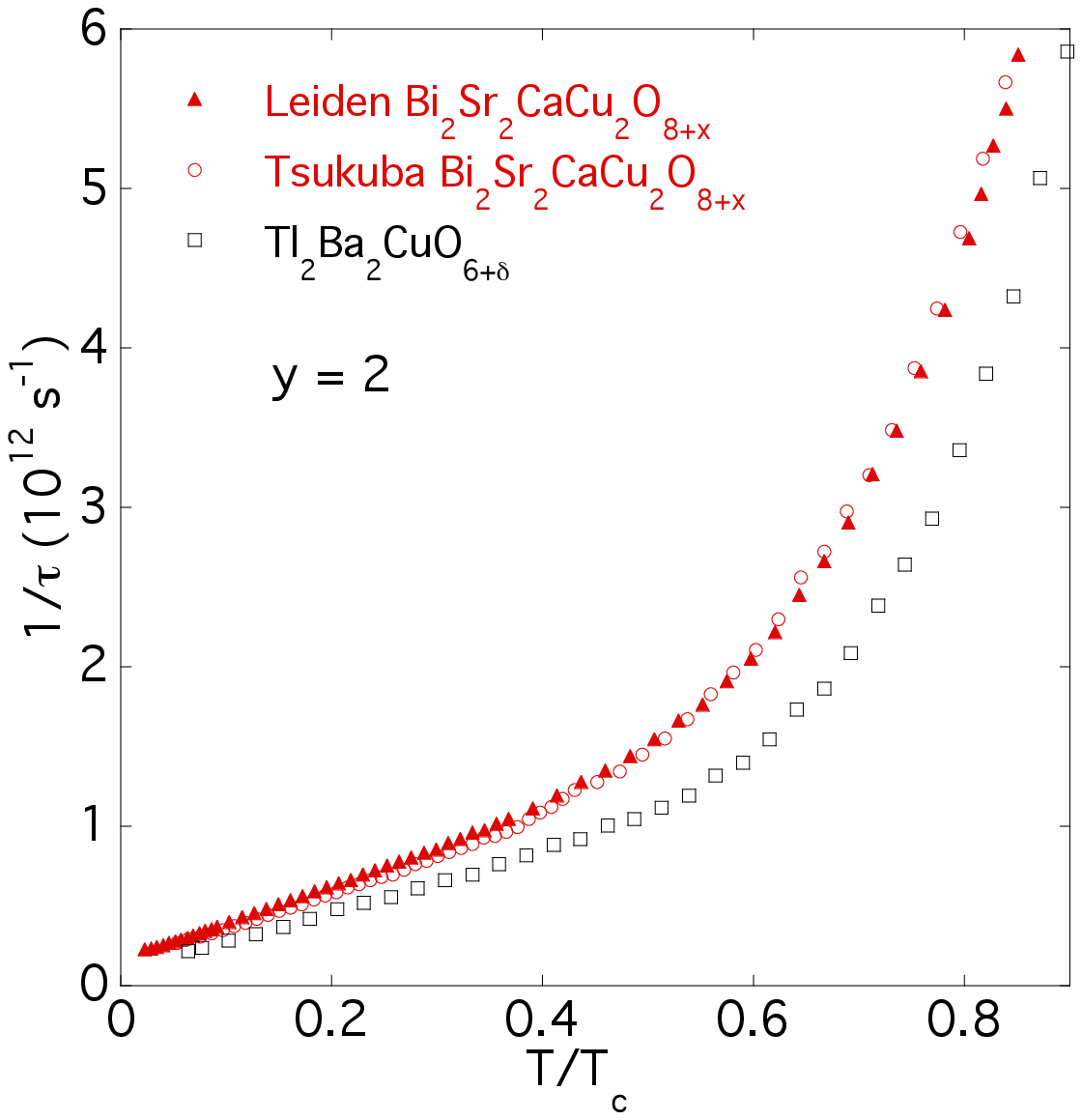}
\caption{(Color online)  Relaxation rates $1/\tau(T)$ of Leiden \bscco, Tsukuba \bscco\ and \tbcod, obtained from a two-fluid analysis using a Drude ($y = 2$) quasiparticle conductivity spectrum.  The low temperature linear regime and the strong temperature dependence of $1/\tau(T)$ at higher temperatures are generic features of the three samples.}
\label{figure5} 
\end{figure}
\begin{figure}[t]
\includegraphics[width = 85mm]{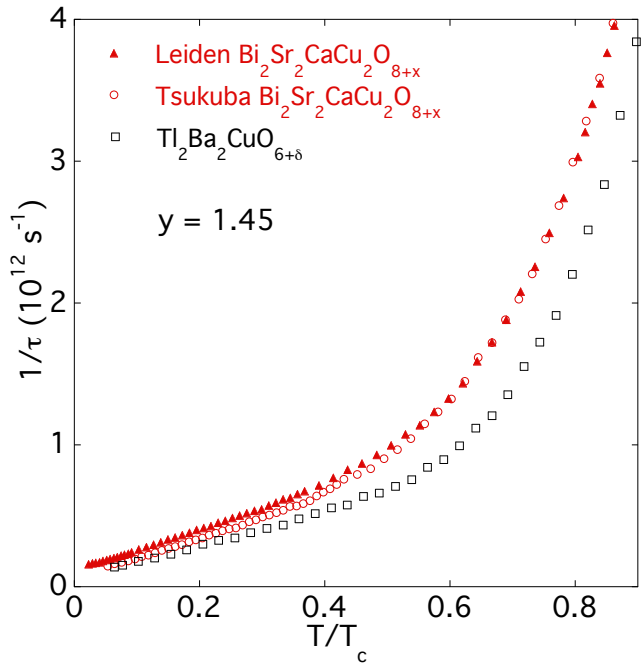}
\caption{(Color online)  Relaxation rates $1/\tau(T)$ of Leiden \bscco, Tsukuba \bscco\ and \tbcod, obtained from a two-fluid analysis using a non-Drude  ($y = 1.45$) quasiparticle conductivity spectrum.  
$1/\tau(T)$ is reduced by a factor of $\approx 1.5$ but is qualitatively the same as for the Drude-limit analysis in Fig.~\ref{figure5}.}
\label{figure6}
\end{figure}

We begin by applying the two-fluid analysis to the new conductivity data from the Leiden \bscco\ crystal.  As mentioned above, the total conductivity spectral weight, parameterized by $\sigma_0$, plays an important role in the analysis.  We demonstrate this as follows.  First, we assume that all conductivity spectral weight condenses into the superfluid as $T \rightarrow 0$ and set $\sigma_0$ by extrapolating $\sigma_2(T)$ to $T = 0$. In this case we get $\sigma_0 = 7.5\times10^7~\Omega^{-1}$m$^{-1}$. On carrying out the two-fluid analysis for $y=2$ we obtain the data shown as open triangles in the plot of superfluid fraction $f_\textrm{s}$ in Fig.~\ref{twofluid}. $f_\textrm{s}(T)$ increases monotonically with decreasing temperature down to 10~K, then peaks and shows a rapid downturn.  The origin of this unphysical behaviour is the assumption that all spectral weight condenses, which can be relaxed by choosing a larger value of $\sigma_0$, corresponding to an increase in the \textit{total} conductivity spectral weight.  There is some uncertainty in how to choose $\sigma_0$, and in the analyses that follow we  set $\sigma_0$ to the smallest value that results in a monotonic temperature dependence of $f_\textrm{s}(T)$.  Other choices of $\sigma_0$ do not affect our conclusions and make no qualitative changes to the form of $1/\tau(T)$.  For the Leiden \bscco\ crystal we have chosen $\sigma_0 = 8.3\times10^7~\Omega^{-1}$m$^{-1}$.  The effect on $f_\textrm{s}(T)$, which is given by the closed symbols in Fig.~\ref{twofluid}, is quite dramatic at low temperatures:  $f_\textrm{s}$ is now monotonic in temperature.  An extrapolation of $f_\textrm{s}(T)$ to $T =  0$ gives a residual, uncondensed spectral weight equal to 11\% of the total (see Table~1).  All microwave and terahertz conductivity
measurements to date find residual conductivity spectral weight:  in \ybco, where the spectrum can be measured in great detail, the residual conductivity spectral weight ranges from 0.5\% to 3\%.\cite{turner03,harris05}  The higher value in \bscco\ is likely due to increased levels of static disorder.  Theorists have had some success in modeling the uncondensed spectral weight using impurities with scattering phase shifts intermediate between the Born and unitarity limits.\cite{carbotte02,carbotte03}

The effect on $1/\tau(T)$ of varying $\sigma_0$ is not nearly as pronounced as it is on $f_\textrm{s}(T)$.  Fig.~\ref{scatteringrate} shows $1/\tau(T)$ inferred from the Leiden \bscco\ data using the same choices of $\sigma_0$ as above.  Increasing $\sigma_0$ gives a small, approximately temperature independent shift of $1/\tau(T)$.  These data also reveal one of the main findings of this paper, the strong linear temperature dependence of relaxation rate at low temperature, which is the behaviour expected for weak-limit impurity scattering in a \mbox{$d$-wave} superconductor with line nodes.  For the physically relevant case in which not all spectral weight condenses, $1/\tau(T)$ approaches a finite low-temperature intercept.  This is also what would be expected in the weak scattering limit, in which the energy dependent relaxation rate is determined by Fermi's Golden Rule and should reflect the presence of a residual density of states $N(E \rightarrow 0)$ induced by pair breaking.  Both the linear temperature dependence of $1/\tau$ and a finite residual relaxation rate are also observed in Ortho-II YBa$_2$Cu$_3$O$_{6.52}$.\cite{turner03}

\begin{figure} [t]
\includegraphics[width = 85mm]{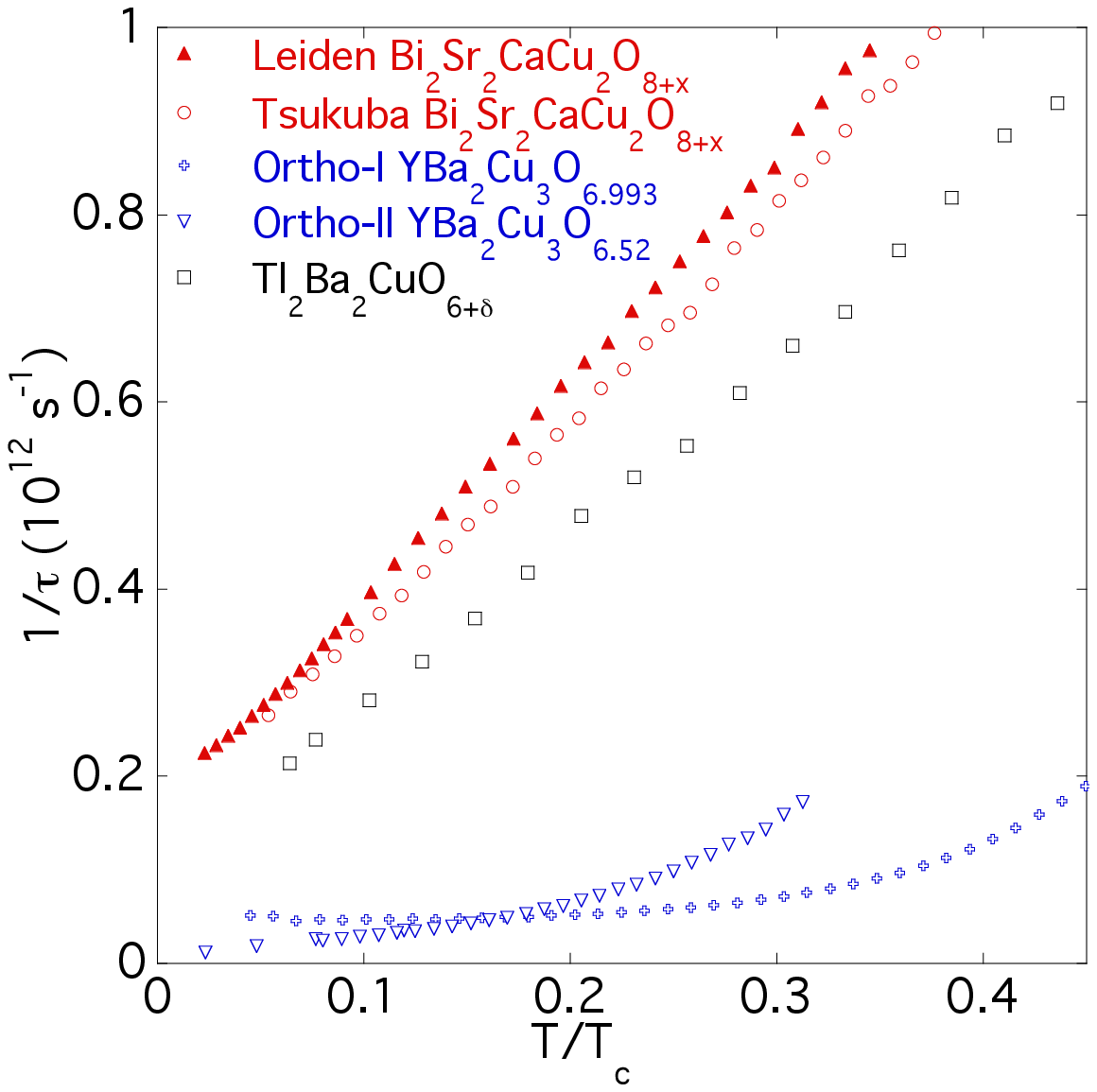}
\caption{(Color online) A close-up view of the low temperature region of Fig.~\ref{figure5}, with the \bscco\ and \tbcod\ data plotted alongside relaxation rate data from microwave spectroscopy of Ortho-I YBa$_2$Cu$_3$O$_{6.993}$ and Ortho-II YBa$_2$Cu$_3$O$_{6.52}$.\cite{hosseini99,turner03,harris05}  Ortho-II YBa$_2$Cu$_3$O$_{6.52}$ also has a linear temperature dependence of $1/\tau(T)$ a low temperature.  The rapid onset of inelastic scattering at higher temperatures is a common property of all the materials shown.}
\label{figure7}
\end{figure}

We now apply the same analysis to our previously published conductivity data on \tbcod\ (\tc~$ = 78$~K) at 35.9~GHz and Tsukuba-grown \bscco\ (\tc~$ = 91$~K) at 34.7~GHz.  To demonstrate the insensitivity of the analysis to the detailed shape of the conductivity spectrum, we carry it out in the $y=2$ (Drude) limit  and for $y= 1.45$, corresponding to the spectral shape observed in Ortho-II YBa$_2$Cu$_3$O$_{6.52}$.  In all cases the two-fluid analysis reveals the presence of significant residual normal fluid at $T = 0$.  We do not include plots of $f_\textrm{s}(T)$, which are similar to Fig.~\ref{twofluid}, but summarize the results for the residual normal fluid fraction in Table~\ref{table1}.

\begin{table}[h]
\begin{center}
 \renewcommand{\arraystretch}{1.5}
\begin{tabular}{|c|c|c|c|}
\hline
 & \raisebox{-0.5ex}{Leiden} & \raisebox{-0.5ex}{Tsukuba}  & \\
 \raisebox{1.7ex}{$y$} & \raisebox{0.5ex}{\bscco} & \raisebox{0.5ex}{\bscco} & \raisebox{1.7ex}{\tbcod} \\
 \hline
1.45 & 0.16 & 0.15 & 0.21 \\
  \hline
2  &  0.11 & 0.10 & 0.15 \\
  \hline
\end{tabular}
\renewcommand{\arraystretch}{1.0}
\end{center}
\caption{Residual normal fluid fraction, $1 - f_\textrm{s}(T\rightarrow 0)$.}
\label{table1}
\end{table}

The temperature dependent relaxation rates for all three samples are plotted in Figs.~\ref{figure5} and \ref{figure6} for $y=2$ and $y=1.45$ respectively.  A close-up of the low temperature region of the $y=2$ data, including a comparison with \ybco\ data, is given in Fig.~\ref{figure7}.  The first noticeable feature is that the plots for different $y$ are qualitatively the same, differing only by a temperature independent scale factor $\approx 1.5$.  This insensitivity of the spectral width parameter, $1/\tau(T)$, to the detailed shape of the conductivity spectrum has its origin in the oscillator strength sum rule.  The superfluid density, which is obtained in our experiment through the simultaneous measurement of $\sigma_2(T)$, fixes the oscillator strength of the conductivity spectrum at each temperature.  A measurement of the \textit{height} of the conductivity spectrum, $\sigma_1(\omega \rightarrow 0)$, then allows a very good estimate of the \textit{width} of the spectrum, $1/\tau(T)$.  The second aspect of the $1/\tau(T)$ data we notice is that there are broadly two regimes of temperature-dependent behaviour.  At low temperatures $1/\tau(T)$ is approximately linear in temperature.  As discussed earlier, this is the expected behaviour of a \mbox{$d$-wave} superconductor in the weak-scattering limit.  Unlike a normal metal, in which $1/\tau(T)$ approaches a constant low-temperature limit due to elastic impurity scattering, elastic scattering in a \mbox{$d$-wave} superconductor is temperature dependent: the degree of elastic scattering is better characterized by the \textit{slope} of $1/\tau(T)$.  At higher temperatures $1/\tau(T)$ grows much faster than linearly with temperature.  We take this to be the onset of \textit{inelastic} scattering.  
\begin{figure} [t]
\includegraphics[width = 85mm]{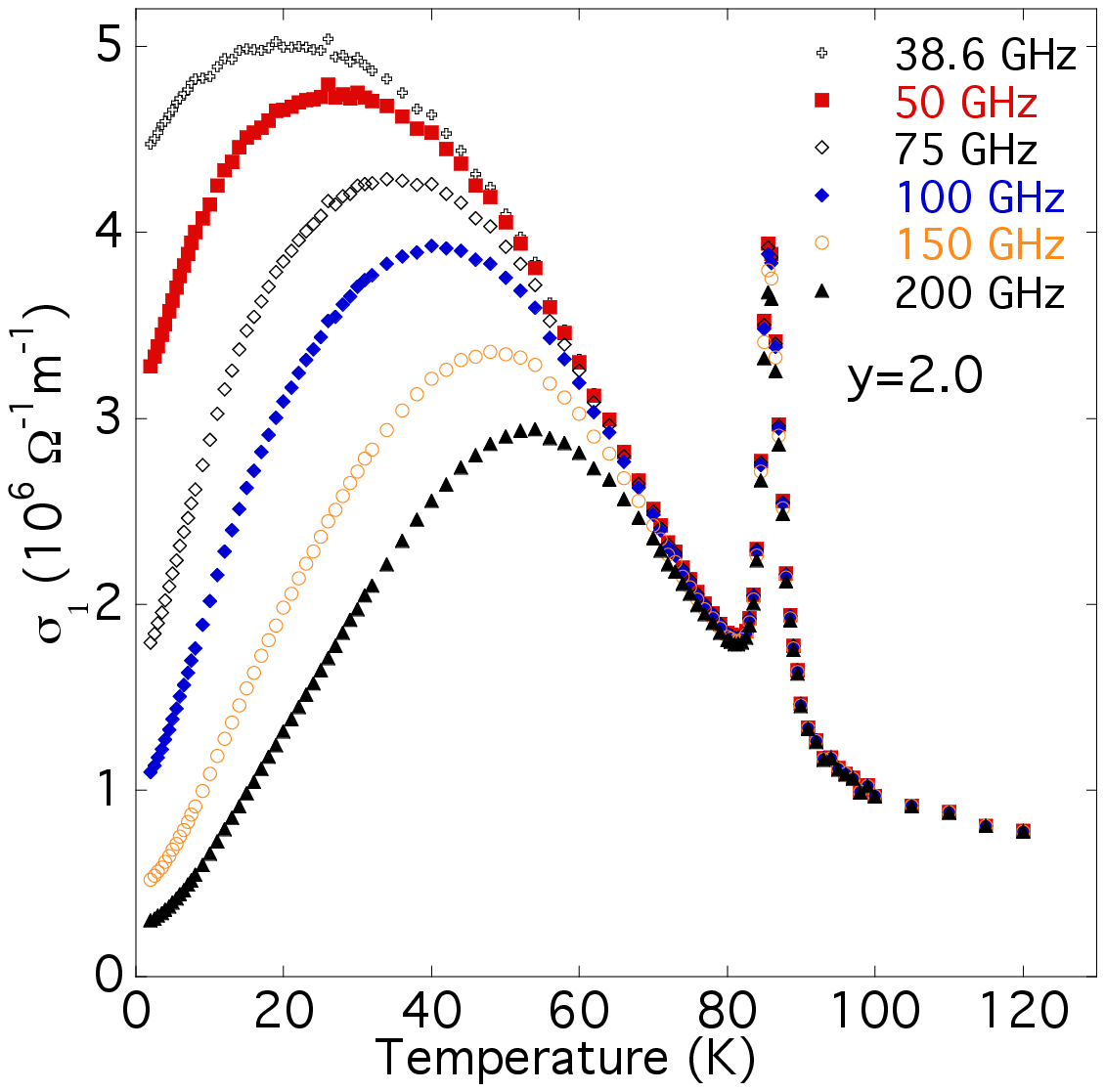}
\caption{(Color online)  Extrapolation of the 38.6~GHz microwave conductivity data from the Leiden \bscco\ crystal to high frequencies, using the two-fluid parameters from the Drude ($y = 2$) analysis.  The extrapolated data are similar in form to terahertz conductivity data on \bscco\ thin films,\cite{corson00} which show clear signs of residual conductivity spectral weight in the zero-temperature limit.}
\label{orenstein2} 
\end{figure}
\begin{figure} [t]
\includegraphics[width = 85mm]{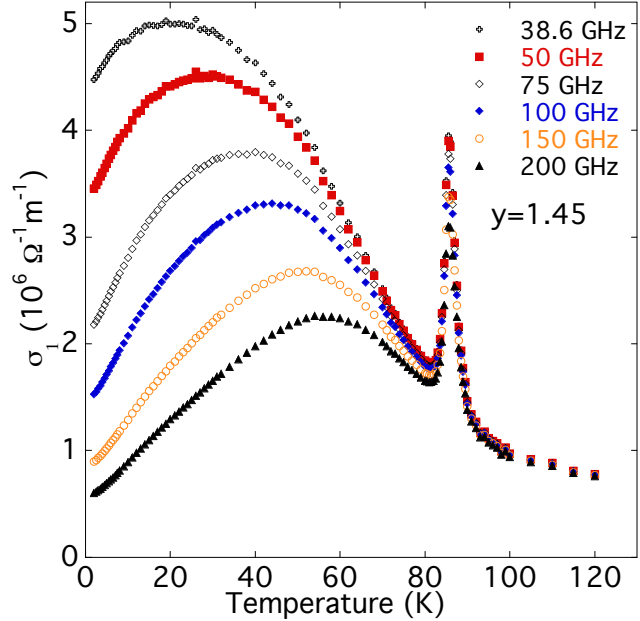}
\caption{(Color online) Extrapolation of the 38.6~GHz conductivity data to high frequencies using two-fluid parameters from the $y=1.45$ analysis.  The qualitative form of the high frequency extrapolation is insensitive to assumptions about the detailed shaped of the quasiparticle conductivity spectrum, making this a robust analysis.}
\label{orenstein1.45} 
\end{figure}
A regime of weak elastic scattering followed by a rapid onset of inelastic scattering was first observed in Ortho-I YBa$_2$Cu$_3$O$_{6.993}$.\cite{hosseini99} In that context Walker and Smith pointed out the importance of Umklapp processes in relaxing electrical currents, showing that in a \mbox{$d$-wave} superconductor such processes become gapped in the superconducting state, leading to an activated exponential temperature dependence of the relaxation rate.\cite{walker00}  The distinction between Umklapp and normal processes has since been included in calculations of quasiparticles scattering from antiferromagnetic spin fluctuations to obtain good agreement with thermal and microwave conductivity data.\cite{duffy01,scalapino05}  The third feature of the data is the similarity of the magnitude of $1/\tau(T)$ in the three materials, being  nearly identical in the two \bscco\ samples and only 30\% smaller in \tbcod.  This suggests an intrinsic origin of the scattering, and it seems reasonable to look to cation nonstoichiometry as the dominant source of disorder.  In both \bscco\ and \tbcod, deviations from cation stoichiometry at the 10\% level are a result of thermal equilibrium fluctuations in composition at the crystal growth temperature.  (Cation nonstoichiometry in \ybco, by contrast, is at the $10^{-4}$ level, due to the high chemical stability of the \ybco\ structure.)  One consequence of the cation nonstoichiometry is the creation of disorder located \textit{off} the CuO$_2$ planes.  This smooths the disorder potentials experienced by electron quasiparticles, which then give rise to predominantly small angle scattering.  Recent work to include the effect of small-angle scattering in the SCTMA theory has been very successful in describing experimental data and represents a major advance in our understanding of the transport properties of the cuprates.\cite{durst00,nunner05}

We finish the analysis and discussion by making a connection with terahertz  experiments on \bscco\ from the Berkeley group.\cite{corson00}  The terahertz measurements span a wide enough frequency range that within the superconducting state they capture most of the low-frequency conductivity spectrum. These experiments were carried out on \bscco\ thin films, so the samples are in principle quite different from the single crystals used in the microwave experiments.  Nevertheless, if a universal picture of quasiparticle scattering dynamics in the cuprates is to emerge, it should be possible to understand both the microwave and terahertz data within a common framework of intrinsic inelastic scattering and sample-dependent disorder.  To motivate that comparison we use the two-fluid parameters for the Leiden \bscco\ crystal to extrapolate the microwave conductivity $\sigma_1(\omega,T)$ into a range of frequencies comparable to those of the terahertz experiment.  The extrapolated microwave conductivity is plotted in Figs.~\ref{orenstein2} and \ref{orenstein1.45}, for $y=2$ and $y=1.45$ respectively.  The frequencies used for the extrapolation are $f = $~50, 75, 100, 150 and 200~GHz.  These are a factor of 4 smaller than the frequencies used in the terahertz experiment, $f = $~0.2, 0.3, 0.4, 0.6 and 0.8~THz, a choice that reflects the lower level of scattering in the single crystals compared to the thin films.  The extrapolated microwave data are qualitatively similar to Fig.~1 of Ref.~\onlinecite{corson00}, with a large residual conductivity $\sigma_1(T \rightarrow 0)$ at low frequencies that becomes small at the higher frequencies.  The ability of Nunner \textit{et al.} to fit our \bscco\ microwave data within the SCTMA framework using a combination of dilute strong scatterers and a high density of weak-scattering extended defects\cite{nunner05} suggests that a similar calculation might account for the terahertz data.

\section{Conclusions}

We have carried out a two-fluid analysis of high resolution microwave conductivity data on \bscco\  and \tbcod\  single crystals.  We have carefully tested the sensitivity of  our analysis to the assumed form of the quasiparticle conductivity spectrum, and find the procedure to be robust. 
In all the \bscco\  and \tbcod\  samples under investigation we observe a rapid collapse of the transport relaxation rate upon entering the superconducting state, in apparent contradiction of photoemission and terahertz spectroscopies.\cite{valla99,valla00,corson00}  We also report the observation of a linear temperature dependence of the relaxation rate, theoretically predicted for \mbox{$d$-wave} superconductors in the weak-scattering limit.  $1/\tau(T)$ increases more rapidly at higher temperatures in a way that is reminiscent of the strong onset of inelastic scattering in \ybco.  On extrapolating the two-fluid model to frequencies up to five times the measurement frequency the extrapolation predicts qualitatively similar behavior to terahertz conductivity data on \bscco\ thin films.\cite{corson00}  Our main conclusions are as follows. Relaxation rates in \bscco\ and \tbcod\ single crystals are substantially higher than in \ybco, although not as high as in \bscco\ thin films.  The form of the scattering, in particular the linear temperature dependence of $1/\tau$, suggests that the dominant elastic scatterers act in the Born limit --- these can likely be attributed to cation nonstoichiometry that creates defects located \textit{away} from the CuO$_2$ planes.  With the linear, low temperature term in $1/\tau(T)$ attributed to elastic scatterers, the remaining temperature dependence of $1/\tau$ in  \bscco\  and \tbcod\  bears a strong similarity to that of \ybco,  pointing toward a universal mechanism for \textit{inelastic} scattering in the cuprates.
\section*{ACKNOWLEDGEMENTS}

We wish to acknowledge useful discussions with D.~A.~Bonn, J.~R.~Cooper, J.~S.~Dodge, W.~N.~Hardy, P.~J.~Hirschfeld, R.~X.~Liang, and T.~S.~Nunner.  This work was supported by the National Science and Engineering Research Council of Canada.  One of us (S.~O.) acknowledges financial support while writing this article from J.~R.~Cooper, through the Engineering and Physical Sciences Research Council of the U.~K.  


\end{document}